# Negative substrate bias induced modifications of the physical properties of DC sputter deposited nano-crystalline Mo thin films

Shilpam Sharma[1*] Abhirami S[1,2], E. P. Amaladass[1,2], and Awadhesh Mani[1,2]

[1]Condensed Matter Physics Division, Materials Science Group, Indira Gandhi Centre for Atomic Research, Kalpakkam 603102, India
[2]Homi Bhabha National Institute, IGCAR, Kalpakkam 603102

The negative bias on the substrate during DC sputter deposition of nano-crystalline molybdenum thin films has been utilized to tune their properties for solar cells and the cryogenic radiation detector applications. Films have been deposited on Si substrates under different out of plane negative biases aiming to improve their physical properties such as the sheet resistance, superconducting transition temperature, width of transition and surface roughness. Significant modifications in the electrical and surface morphological properties of nano-crystalline Mo thin films have been reported. The superconducting transition temperature and crystallite size of the films does not show much improvement but the surface roughness and electrical resistivity of the film have been improved by the application of substrate bias.

# 1. Introduction

Thin films of Molybdenum (Mo) form the backbone of several of the modern technological applications. The superior properties like stability at high process temperatures, formation of low resistivity ohmic contacts and insolubility with the normal metals like Cu and Au, make Mo a material of choice for the fabrication of good quality devices. Presently, the Mo thin films are indispensable for applications like superconductor based ultra-high resolution radiation spectroscopy [1-4], $Cu(In_{1-x}Ge_x)Se_2$ (CIGS) thin film solar cells [5-8] and the surface acoustic resonators [9]. While traditionally, Mo films have been used as back electrode in the case of thin film solar cells and resonators, during the recent decades the Mo films in close proximity with normal metals like copper or gold in bilayer or multilayer formations find its use as a sensitive superconducting calorimeter for the radiation spectrometers [1]. They are also extensively used as diffusion barrier in the metal oxide semiconductor based microelectronics industry [10-12]. All these applications require stringent control of the surface, mechanical and electrical properties of Mo thin films deposited on different kinds of substrates.

Being a refractory metal, Mo can be deposited in thin film form using different physical vapor deposition techniques like electron beam evaporation [11, 13], ion beam and magnetron sputtering systems [4]. Among these, the direct current (DC) and RF magnetron sputtering are fast and proficient techniques for the controlled deposition of large surface area coatings and films [14]. During the sputter deposition, various process parameters can be easily manipulated for achieving optimum quality, device grade thin films. For instance, the sputtering power, substrate temperature, target-substrate distance and background gas pressure are some of the many process parameters that can be tuned to achieve desirable properties like low surface roughness, high conductivity, high residual resistivity ratios (RRR) and sharp superconducting transition temperatures ($T_C$).

Among others, the sputtering gas pressure during the growth of the films has been shown to affect the transport and morphological properties of the films [5, 6, 15]. It is well reported that the films deposited under high Ar pressure have porous microstructure with tensile stress and usually adhere well to the substrate [6]. However, the porous micro-structure of the films deposited under high pressure gives rise to poor electrical conductivity [15]. On the contrary, the Mo films deposited under low Ar pressure have compressive stress, poor adhesion but high electrical conductivity. It is for this reason, the Mo back contact layers are generally deposited as graded layers with initial deposition at high pressure and then subsequent layers at reduced pressure thus providing a low resistivity layer that adheres well to the substrate [6]. Likewise, the effect of sputtering power and the resulting DC bias of target in the case of RF sputtering are also studied for the optimization of superconductivity in Mo films [3]. Most of these process parameters alter the deposition rate, surface mobility and energy of adatoms to produce the desired quality films [16]. In our previous report [15], we have shown that the background Ar pressure during the film growth has a significant role in determining the superconducting and normal state properties of the nano-crystalline Mo films. The migration of sluggish Mo adatoms on substrate gets further hindered by the trace oxygen present in the system and thus the films deposited at low sputter rate tend to have nano-crystalline microstructure with low RMS surface roughness. Owing to the size effect in the nano-crystalline Mo films, the superconducting transition temperature ($T_C$) of the films has been shown to increase up to as high as 2-6 K. An increase in the grain size may bring the $T_C$ close to bulk but may increase the surface roughness also.

Thus, in the present study we have attempted to improve upon the superconducting and morphological properties of the nano-crystalline Mo films by applying out of plane negative bias on the substrate during sputtering. The aim was also, to produce nano-crystalline Mo films with low sheet resistance, low RMS roughness and good adhesion with small tensile stress which can be used in place of graded Mo back contacts deposited at varying Ar pressures [6]. The negative bias on substrate is known to significantly alter the physical properties of the films [17, 18]. There are only a few reports on the use of substrate bias for modification of the morphological, transport and superconducting properties of particularly the nano-crystalline Mo films [19, 20]. This kind of study will be helpful in the case of magnetron sputter deposition as a large flux of low energy sputtered ions, electrons and oxygen ions can reach the substrate. The application of substrate bias can increase the energy and/or control the direction of these incident charged species which can have an impact on the properties of growing thin films.

## 2. Experimental Details

Nano crystalline thin films of molybdenum have been deposited on SiO$_2$ coated (300 nm) Si(1 0 0) substrates using a home built DC magnetron deposition system. The deposition system is equipped with three home built 1" DC magnetron cathodes arranged in planetary configuration along with a load lock system for loading and unloading of substrates under ultra-high vacuum. The substrates were thoroughly cleaned by ultra-sonication in boiling acetone followed by ultra-sonication in ethyl alcohol and finally blow dried using dry nitrogen. Prior to deposition, the system was pumped to a residual pressure of ~ 1x10$^{-7}$ mbar using a 150 lit/sec turbo molecular pumping system and sputtering was performed in a background Ar pressure of 3.5 x 10$^{-3}$ mbar. A commercial 1", high purity (99.95%) Mo target (Testbourne ltd. U.K.) was sputtered using UHP (99.9995%) Ar ions. For all the depositions, the substrate to target distance was kept fixed at 110 mm. Before deposition of the films, the target surface was cleaned of residual oxides by pre-sputtering for roughly 5 minutes. During deposition, the substrate was kept at room temperature and the holder was biased negatively up to 150 V in the steps of 50 V using a high voltage DC power supply. The Mo films were deposited for 480 seconds with Mo target sputtered using a constant DC current fixed at 120 mA. During the deposition, the voltage on the cathode ranged from 380 to 390 V which resulted in ~46 W of DC power delivered to the target. An average deposition rate of 0.8 Å/sec was achieved. The phase purity of the films has been confirmed using grazing incidence x-ray diffraction (GIXRD) measurements performed using Cu Ka radiation (Inel make Equinox 2000 X-ray diffractometer). The film thickness was estimated using the Dektak surface profilometer (Bruker) and the surface morphology was studied using scanning tunneling microscope (STM) (Quazar Technologies Pvt. Ltd, India,). The transport measurements up to sub-Kelvin temperatures have been performed in the dilution refrigerator (Oxford Instruments, U. K. make cryofree Triton 200-10). The dilution refrigerator is equipped with shielded coaxial cables which enables measurements of 5 samples in a single run. The transport measurements were performed by the four-point probe method in the van der Pauw configuration using a home built commutator card along with a 20 channel multiplexer (Agilent 34970A). The multiplexer was used to select one of the multiple samples loaded in the cryostat and the commutator card was used to measure the sheet resistance of the selected sample by commuting the current and voltage leads in different van der Pauw configurations.

## 3. Results and Discussion

GIXRD measurements have been performed to characterize the phase purity of Mo thin films. The plots of intensity as a function of diffraction angle are presented in figure 1. Reflections originating only from the Mo *d*-spacings could be seen in the plots which indicate the phase purity of the samples. Three main peaks with (1 1 0), (2 0 0) and (2 1 1) orientations could be observed in the GIXRD plot. The GIXRD data has been collected for all the samples for 40 minutes using a curved position sensitive detector while the angle of incidence was kept fixed at 2°. The peak intensity shows a dependence on the substrate bias. It can be noticed from figure 1b that the intensity of the (1 1 0) peak decreases as the negative substrate bias increases from 0 to -150 V. The intensity of (2 0 0) peak decreases and the peak has not been observed in the sample deposited at -150 V. The FWHM of the (1 1 0) peak plotted in figure 1c has also been observed to decreases with the substrate bias. The FWHM for the film deposited without bias has been found to be 0.98 degrees and it decreases monotonically to 0.82 degrees for the sample deposited at -150 V.

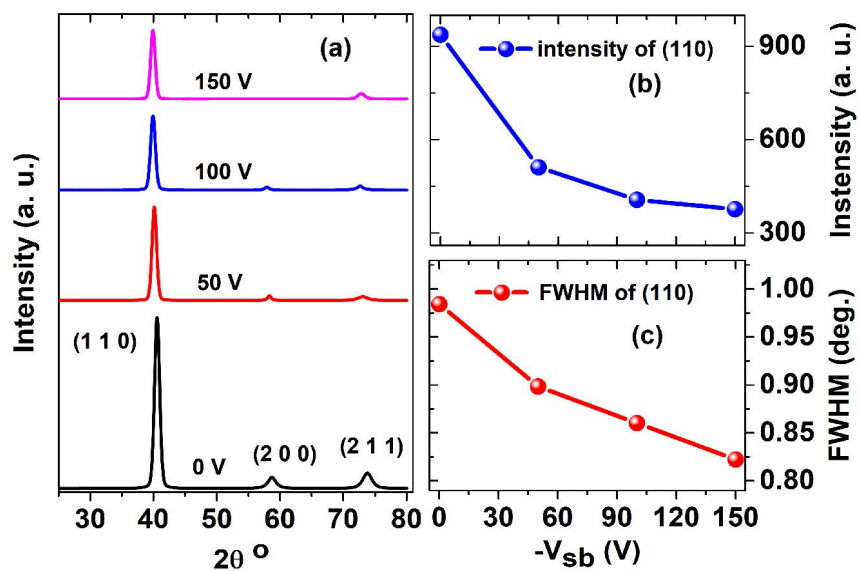

**Figure1a:** GIXRD plot of polycrystalline Mo thin films deposited under different negative substrate biases applied during the film deposition. The intensity of the peaks has been found to decrease with the increase in the substrate bias. The shifting of peaks towards lower angles indicates increase in the out of plane tensile stress in the films due to application of substrate bias. **(b), (c):** Intensity and FWHM of (110) peak as a function of substrate bias shows a decrease.

In addition, all the peaks in the GIXRD of the films deposited under substrate bias can be seen to shift towards lower 2θ values. This shift could be attributed to the lattice expansion along the growth direction thereby giving rise to an increase in the out of plane tensile stress in the films [3]. The strain in the films has been estimated by calculating the change in the lattice spacing of the films as compared to the bulk value of 3.1472 Å [21]. The strain has been found to increase monotonously from 0.15% at 0 V substrate bias to 1.5% at -150 V and can be attribute to an increase in the defects [22] due to the application of substrate bias. Similar values of %strain has been reported in several other reports on sputtered Mo thin films [7]. The thickness of the films, estimated using profilometer measurements, has been found to slightly depend on the substrate bias during deposition. The thickness of the film decreased from ~40 nm for the film deposited at zero bias to ~32 nm for the film deposited at -150 V substrate bias. The reduction in the thickness of the films could be due to the re-sputtering of the growing film by bombardment of positive ions at the film surface. This bombardment by highly energetic ions and reflected neutrals can take out loosely bound low energy atoms from the film surface. The observed reduction in the films thickness could be the reason for the decrease in the intensity of the peaks and increase in the tensile strain with increase in the substrate bias.

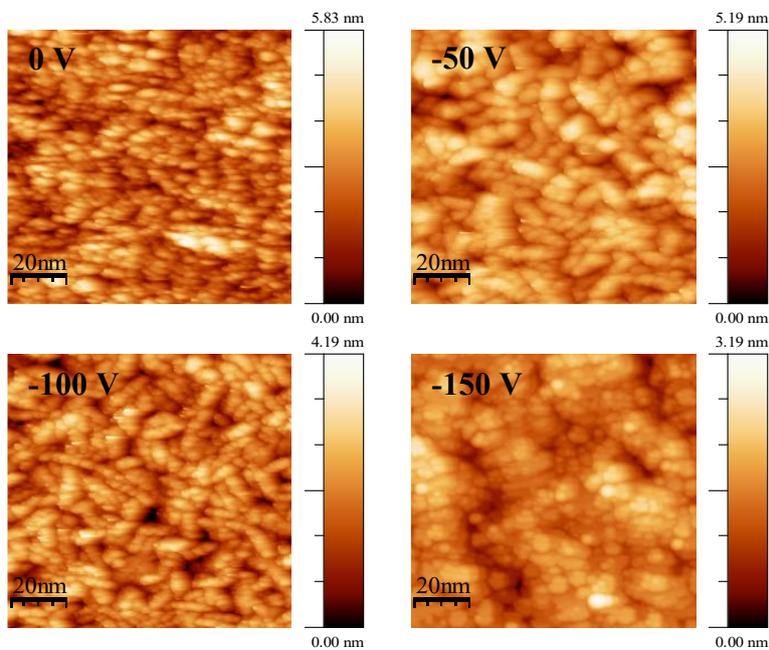

**Figure 2:** The STM images showing the surface topographies of the films deposited under different substrate bias. The morphology of the films shows a dependence on the applied substrate bias. The height of the crystallite also decreases with increase in the substrate bias.

In order to estimate the grain size and RMS surface roughness, the film topography has been recorded using STM with a Pt-Ir tip. The STM nano-graphs of the Mo films deposited under different negative substrate bias are shown in figure 2. The morphology of the films can be seen to get affected due to applied substrate bias. The film deposited without substrate bias has roughly spherical nano-crystallites with sharp boundaries whereas, the films deposited under substrate bias shows a coalescence of the crystallites with obscure boundaries giving rise to elongated grain morphologies. The film deposited at -150 V bias have the most obscure grain boundaries. It can also be noticed from the figure 2 that the maximum height of the nano-crystallites decreases with the substrate bias that has a bearing on the overall thickness of the films which decreases with increase in the substrate bias. Similar morphologies and their modifications have already been reported for the Mo films [19, 23, 24], but the crystallite size of our films is much smaller than those reported earlier.

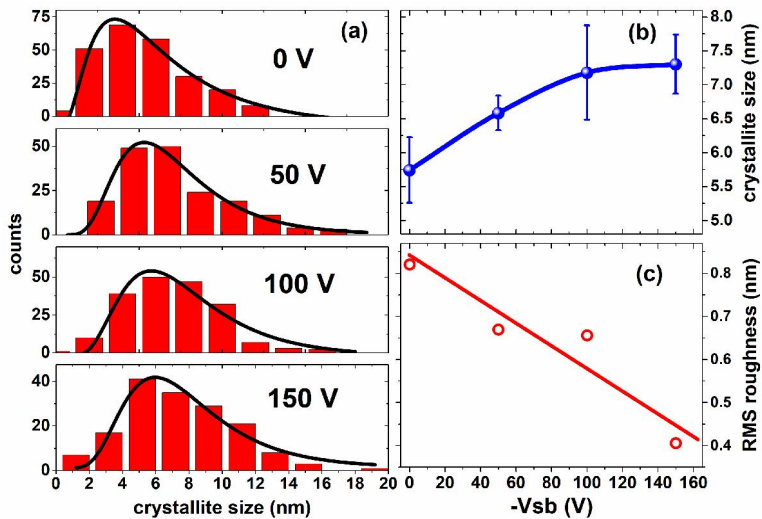

**Figure 3a:** The crystallite size distribution and its fitting to log-normal distribution. **(b):** the plot of average crystallite size as a function of negative substrate bias. The substrate bias only marginally affects the crystallite sizes of the films. **(c):** RMS surface roughness of the films as a function of negative substrate bias. The surface roughness decreases with increase in the substrate bias. The straight line is for visual guidance only.

The average crystallite size of the Mo films has been estimated by analyzing the STM nano-graphs using ImageJ [25] and NT-MDT SPM image processing software. The size distribution of ~100-200 particles has been fitted to the log-normal distribution and the average crystallite size has been calculated. The crystallite size distribution and the log normal fit for all the thin film samples are presented in figure 3a. The films have been found to be nano-crystalline in nature. It has been observed that the average lateral crystallite size increases marginally from ~ 5.7 nm to 7.3 nm for the increase in the substrate bias from 0 to -150 V (cf. fig 3b). This increase in the lateral crystallite sizes is in-line with the decrease in the FWHM of (110) peak. Similar nano-crystalline Mo films have been reported in literature by Vink et al. [22]. The marginal effect of substrate bias on the grain size of Mo films has already been reported by Xu et. al. [17]. The substrate bias induced coalescence of the nano-crystallites may have an impact on the sheet resistance and the RMS surface roughness of the films.

The effect of substrate bias on the RMS roughness of the films has been estimated by analyzing the STM images using the WSXM software [26]. As can be seen from figure 3c, the RMS surface roughness showed a decrease from 0.82 nm to 0.4 nm as the negative bias increased from 0 to 150 V. The roughness of the samples deposited under -50 V and -100 V bias have been observed to be nearly the same. The decrease in the maximum height and coalescence of the nano-crystallites could be the reason for the decrease in the RMS surface roughness of the films. These values of the surface roughness are comparable to the RMS values reported by Fabrega et al. [3] but are fairly low as compared to several other similar studies on the Mo thin films [9, 27, 28]. The low surface roughness is desirable in solar cell applications as it helps in achieving sharp interfaces between the Mo film and subsequent absorber layers thus leading to an ohmic contact with low contact resistance, whereas, in the case of Mo based micro calorimeters the sharp interface between Mo and normal metal layers dictates sharpness of the superconducting transition.

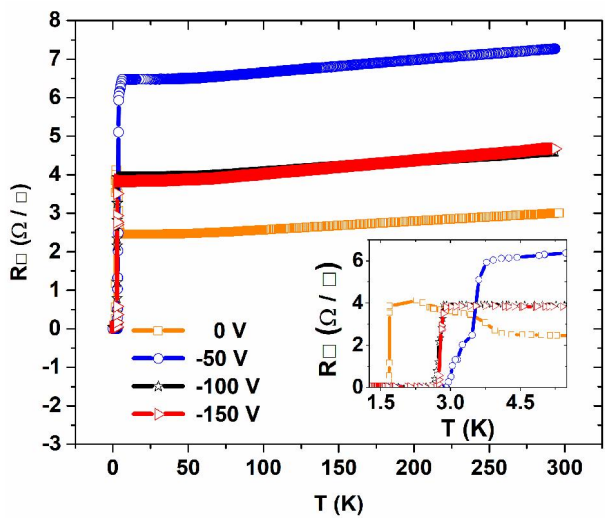

**Figure 4:** Sheet resistance as a function of temperature for the films deposited under substrate bias. The films show metallic R(T) behavior with RRR >1 and a positive coefficient of resistivity. **Inset:** superconducting transition in the films deposited at different substrate biases.

The electrical transport properties of the films were studied using a commercial 10 mK dilution refrigerator. The plots of sheet resistance as a function of temperature up to 10 mK are presented in figure 4. As can be noticed from the inset of figure 4, all the samples have been found to show superconducting transition to zero resistance state at different temperatures much higher than the bulk $T_C$. The films show metallic characteristics with positive coefficient of resistivity but due to the nano-crystallinity of the films, the size effects were dominant [15] and the $T_C$ was found to be higher than the bulk $T_C \sim 920$ mK. A broad peak prior to the superconducting transition can be noticed in the R(T) of the sample deposited without substrate bias. This is a signature of granular film having poorly connected grains embedded in a matrix that is either non-superconducting or superconduct at a lower $T_C$ [29]. The R(T) plots of the samples deposited under negative bias does not show peak prior to the superconducting transition thus pointing to better grain connectivity due to their coalescence as observed in STM micrographs in figure 2. The $T_C$ onset in film deposited under -50 V bias roughly matches with the temperature at which R(T) peak starts in sample deposited without bias (c.f. inset of fig. 4). So it is possible that the grains in the zero bias sample also become superconducting at ~ 4K but due to granular nature of the film, bulk $T_C$ could be seen at a much lower temperature, whereas the sample deposited under -50 V start showing fall in R(T) at ~3.7 K itself due to better grain-grain contacts. For the samples deposited under substrate bias, the $T_C$ correlated well with the average crystallite size (c.f. fig. 3b). It can be observed to decrease from 3.7 K for -50 V substrate bias to 2.7- 2.8 K for higher substrate biases (c.f. fig. 5). RRR ($R_{300K}/R_{4K}$) plotted in figure 5 also showed similar dependence on the substrate bias and it decreased initially from 1.23 to 1.12 for 0 V to -50 V but then recovered to 1.20 at -150 V bias. The room temperature sheet resistance of the films was found to increase from 2.5 Ω/□ to 7.2 Ω/□ at -50 V which subsequently decreases to 4.6 Ω/□ at -150 V bias. The sheet resistance values of our Mo thin films compares well with the reported values [17, 23]. The room temperature resistivity ($\rho_{300}$) of the films has been estimated using the sheet resistance and film thickness. The $\rho_{300}$ of the films has been found to be higher than the bulk value of 5.4 μΩ-cm. It has been estimated to be varying between 10-14 μΩ-cm for samples deposited at lowest and highest biases, but goes to a high of 28 μΩ-cm for -50 V substrate bias.

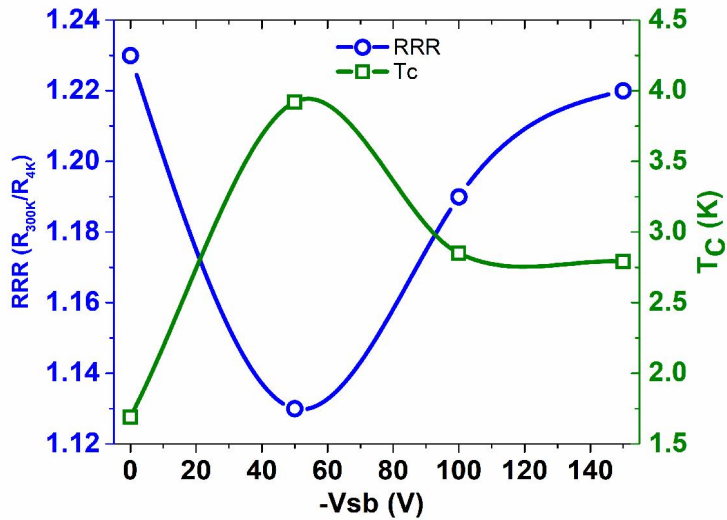

**Figure 5:** Residual resistivity ratio and the superconducting transition temperature as a function of negative substrate bias. The RRR and $T_C$ correlates well among each other. The $T_C$ is seen to be higher than the bulk due to finite size effect of the nano-crystalline samples.

Slightly higher value of the resistivity (~32 μΩ-cm) have been observed in our earlier work [15] where Mo films were deposited at similar Ar background pressure but using a much lower sputtering power. The resistivity of our films is much lesser than the values reported (~ 40-80 μΩ-cm) for the Mo films with comparable or larger crystallite sizes [8, 27, 30]. The effect of oxygen contamination during the film growth upon its electrical properties is well documented [12, 19, 20]. The reduced sheet resistance and resistivity of the films in present work could be attributed to lesser oxygen contamination due to increase in the deposition rate and repulsion of negative oxygen ions away from the growing films [20]. The resistivity of the films depends upon the film microstructure and the impurity content of the film. Both of these get affected by the amount of oxygen present at the time of deposition. For the case of sedentary refractory materials with low surface mobility, the growth of fibrous grains decreases with the increase in the oxygen content. The oxygen hinders the grain growth and remains as oxide in the grain boundaries thus giving rise to large electron scattering at the grain boundaries. Thus, reduced hindrance to the grain growth due to depletion of oxygen ions in the vicinity of the substrate along with increased mobility of Mo adatoms might have helped the grains in our samples to grow larger as compared to the film deposited without bias. Under the influence of negative bias at the substrate, the positive ions from the plasma bombard the surface of growing film. This, in addition to the re-sputtering of loosely bound adatoms and adsorbates from the film surface, promotes the atomic rearrangements by increasing the energy and mobility of adatoms [16]. The end effect of application of substrate bias is usually a thin film with dense microstructure with less oxygen content in the grain boundaries. This seems plausible in our Mo films too as all the films have passed the tape test without getting peeled off thus pointing towards a dense film with good adhesion.

## 4. Conclusion

In conclusion, the attempt to improve the morphological and electrical transport properties of DC sputter deposited nano-crystalline Mo films by employing the substrate bias technique has resulted in the growth of films with very low surface roughness, good adhesion and low resistivity. The Mo film deposited without substrate bias has been found to be granular as indicated by the peak in R(T) prior to superconducting transition. The absence of such peak in the R(T) data of samples deposited under substrate bias points towards better crystallinity in these films. The films are seen to exhibit tensile stress in the direction of the growth which is found to increase with the substrate bias. The films have been found to be nano-crystalline with marginal improvement in their lateral crystallite size. The dense microstructure of the films has resulted in the better film adhesion. While the crystallite size, normal state electrical and superconducting properties of the films like RRR and $T_C$ does not show much improvement upon application of substrate bias, nonetheless they do correlate well among each other. The nano-crystalline nature of the films gives rise to enhancement of $T_C$ due to the finite size effects. The conductivity of the films has improved due to the application of bias, which could be an effect of the $O^-$ ions repulsion during film growth and increased surface mobility of the adatoms. Even though the superconducting $T_C$ of the films is higher than the bulk $T_C$ even after application of the substrate bias, the reduction in the surface roughness and sheet resistance with the increase in the negative substrate bias can be utilized for making device grade Mo back electrodes in solar cells and resonator applications.

**Acknowledgements**

The authors would like to thank P. Magudapathy, Materials Science Group, IGCAR for GIXRD measurements on the Mo thin films.